\documentclass[a4paper,aps,pra,twocolumn,twoside,nofootinbib]{revtex4-1}

\usepackage{amsmath}
\usepackage{amsfonts}
\usepackage{amssymb}
\usepackage{dsfont}
\usepackage{paralist}
\usepackage{graphicx}
\usepackage{hyperref}
\usepackage{bm}

\newcommand{\mbf}{\bm}

\newcommand{\doubleover}[1]{\overline{\overline{#1}}}
\newcommand{\dd}[1]{\doubleover{#1}}
\newcommand{\ddind}[2]{\dd{#1}{}^{#2}}
\newcommand{\textindex}[1]{\mbox{\tiny #1}}

\begin{document}
\title{The non-birefringent limit of all linear, skewonless media and its unique light-cone structure}
\date{\today}

\author{Alberto Favaro}
\affiliation{Department of Physics, Imperial College London, Prince Consort Road, SW7 2AZ, United Kingdom}
\email{alberto.favaro04@imperial.ac.uk}
\author{Luzi Bergamin}
\affiliation{Department of Radio Science and Engineering, Aalto University, School of Science and Technology, P.O. Box 13000, 00076 Aalto, Finland}
\email{luzi.bergamin@kbp.ch}

\begin{abstract}
Based on a recent work by Schuller et al., a geometric representation of all skewonless, non-birefringent linear media is obtained. The derived constitutive law is based on a ``core'', encoding the optical metric up to a constant. All further corrections are provided by two (anti-)selfdual bivectors, and an ``axion''. The bivectors are found to vanish if the optical metric has signature (3,1) -- that is, if the Fresnel equation is hyperbolic. We propose applications of this result in the context of transformation optics and premetric electrodynamics.
\end{abstract}

\maketitle

\section{Introduction}
In recent years there has been an increasing effort towards understanding and developing non-birefringent, exotic media. The motivation behind this trend appears to be twofold; on the one hand, many modern metamaterials -- particularly in the context of transformation optics \cite{Pendry:2006Sc,Leonhardt:2006Nj} -- are (bi-)anisotropic and yet strictly non-birefringent. On the other hand, the derivation of non-birefringent constitutive relations is a cutting-edge topic in the premetric electrodynamics of spacetime \cite{Hehl:2003, Lammerzahl:2004Rl, Itin:2005Nb}.

Correspondingly, this paper finds its first application in the design of artificial media whose optical response is independent of the field polarization. Most notably, we derive a simple, relativistic constitutive law which parametrizes all non-birefringent, skewonless materials in terms of a symmetric matrix, two (anti-)selfdual bivectors and an axion\footnote{A ``skewon'' arises, for example, when the permittivity tensor $\varepsilon^{ij}$ or the permeability tensor $\mu_{ij}$ are not symmetric. An ``axion'', instead, contributes to the medium response with a non-reciprocal, isotropic term. See Section \ref{sec:2} for further details.}. Conveniently, the pivotal symmetric matrix is proportional to the optical metric, while the (anti-)selfdual bivectors vanish if the Fresnel equation is hyperbolic -- that is, Lorentzian (3,1). By virtue of this result we also investigate on which transformation media can be attained whilst avoiding birefringence.
Remarkably, it emerges that the key choice is to select vacuum as the space on to which to perform the ray-guiding coordinate change\footnote{The vacuum ansatz can correspond to a curved spacetime -- provided the appropriate refinements are considered \cite{Leonhardt:2009Ec}.}. Other options are recognized to be viable. However, they require introducing an axion and/or a metric signature different from $(3,1)$. 
\newline
\newline
A second application of this paper targets the empty space structure of electrodynamics, rather than exotic materials. The seminal works on this topic -- due to Hehl, Obukhov and Rubilar \cite{Hehl:2003,Hehl:2005hu, Rubilar:2002Le} -- demonstrated that the light-cone structure of spacetime can be re-derived using some minimal, premetric, experimentally justified assumptions. In particular, one obtains the conventional vacuum response merely by setting the skewon part to zero, while enforcing a specific closure condition. A similar, yet separate, scheme is pursued by  L\"ammerzahl, Hehl \cite{Lammerzahl:2004Rl} and Itin \cite{Itin:2005Nb}. They achieve the customary light-cone by ruling out birefringence and requiring that running-wave solutions must exist in all directions. So far, no link has been made between this latter approach and the former skewonless, closure-abiding one. Our work presents a direct connection, at least for media with no skewon. 

This paper is organized as follows: firstly we introduce bi-anisotropic local linear materials, the Fresnel equation and the appropriate relativistic toolbox (Section \ref{sec:2}). Subsequently, a classification of all non-birefringent media is obtained and the corresponding compact representation is promptly developed (Sections \ref{sec:3A}--\ref{sec:3C}, and the worked example of Section \ref{sec:3D}).  The results are then applied both in the context of premetric electrodynamics (Section \ref{sec:4A}) and in the context of transformation optics (Section \ref{sec:4B}). We draw our conclusions in Section \ref{sec:5}. The Appendices \ref{sec:appA}--\ref{sec:appC} summarize some technical aspects, mostly concerning the work of Ref.\ \cite{Schuller:2010Ap}.

\section{Fresnel equation and non-birefringence of linear media}
\label{sec:2}
Since this paper is partly targeted to applications in engineering, we assume the existence of a ``background'' metric $\dd g$, thus confining all premetric refinements to Section \ref{sec:4A}. We consider linear, non-dissipative media, for which we introduce a relativistic (frame independent) representation. Suitable starting points are the canonical Boys-Post relations
\begin{align}\label{boyspost}
 \mbf{ D} &=\dd \varepsilon \cdot \mbf E +\dd \alpha \cdot \mbf B\ , & \mbf{ H} &=\dd \beta\cdot \mbf E+\ddind \mu{-1}\cdot\mbf B\ ,
\end{align}
which define the permittivity $\dd \varepsilon$, the inverse permeability $\ddind \mu{-1}$ and the magneto-electric couplings $\dd \alpha$ and $\dd \beta$ \cite{Lindell:DiffForms}. More compactly, one can also encode the medium response as:\footnote{The minus sign appearing in front of $\mbf E$ is clearly superfluous -- but only for the purpose of this equation. In many other occasions it is required so as to comply with Lenz's law \cite{Hehl:2003}. }
\begin{equation}
\label{6x6chi}
\begin{bmatrix}
\mbf D\\\mbf H
\end{bmatrix}
= \begin{bmatrix}
             -\dd \varepsilon & \dd \alpha \\ -\dd \beta & \ddind \mu{-1}
 \end{bmatrix}
\begin{bmatrix}
-\mbf E\\\mbf B
\end{bmatrix}\ .
\end{equation}
The 6$\times$6 block matrix, which is identified here, is often labeled $\chi^{IJ}$ -- with $I$ and $J$ varying between 1 and 6. Furthermore, it is common practice to introduce the identifications $(F_{I})=(-\mbf E;\mbf B)$ and $(W^{I})=(\mbf D;\mbf H)$. This allows one to re-express the map \eqref{6x6chi} as:\footnote{Throughout this paper we make use of Einstein's summation convention.}
\begin{equation}\label{const6x6}
W^{I}=\chi^{IJ}F_{J}\ .
\end{equation}
A relativistic notation is now within reach and is achieved by implementing spacetime indices, ranging from 0 to 3 and commonly denoted by Greek letters\footnote{Given a coordinate patch $\{x^\mu\}$, time is parametrized by $x^0$, while space is spanned by the components $x^i$, $i=1,2,3$.}. With this notation in mind, Eq.\ \eqref{const6x6} is immediately translated to
\begin{equation}
W^{\mu\nu}=\frac{1}{2}\chi^{\mu\nu\alpha\beta}F_{\alpha\beta}\ ,
\end{equation}
where the quantities $F_{\alpha\beta}$ and $W^{\alpha\beta}$ are each defined through an appropriate tableau \cite{Post}:
\begin{align}
  \label{e:faraday-tensor}
F_{\alpha\beta}&=
\begin{bmatrix}
      0&-E_{1}&-E_{2}&-E_{3}\\
E_{1}&0&B_{3}&-B_{2}\\
E_{2}&-B_{3}&0&B_{1}\\
E_{3}&B_{2}&-B_{1}&0
    \end{bmatrix}\ ,\\
    \label{excitationtensor}
W^{\alpha\beta}&=
\begin{bmatrix}
      0&D_{1}&D_{2}&D_{3}\\
-D_{1}&0&H_{3}&-H_{2}\\
-D_{2}&-H_{3}&0&H_{1}\\
-D_{3}&H_{2}&-H_{1}&0
    \end{bmatrix}\ .
\end{align}
By inspection -- or by observing that $F_{\alpha\beta}$ and $W^{\alpha\beta}$ are antisymmetric -- one can conclude that each tableau contains only 6 independent entries. These can in turn be collected according to the rule
\begin{equation}
\label{munutoImap}
  \{[01],[02],[03],[23],[31],[12]\}\rightarrow\{I=1,2,\dots,6\}\ ,
\end{equation}
so as to retrieve the 6-dimensional vectors $F_{I}$ and $W^{I}$. Using an analogous argument, since
\begin{equation}
\chi^{\mu\nu\alpha\beta}=-\chi^{\nu\mu\alpha\beta}=-\chi^{\mu\nu\beta\alpha},
\end{equation}
one can always switch between the spacetime form $\chi^{\mu\nu\alpha\beta}$ and the 6$\times$6 form $\dd \chi=\{\chi^{IJ}\}$\ \footnote{Additional information about this topic can be found in \cite{Hehl:2003,Post}. In all that follows, we use the relativistic and 6$\times$6 notations interchangeably.}. As one might expect, the relativistic representation is particularly convenient when considering a change of frame. More specifically, given an arbitrary non-singular transformation matrix $L^{\alpha'}_{\ \alpha}$, one has that (cf.\ Sec.\ \ref{sec:4A})
\begin{equation}\label{chiframetransform}
  \chi^{\mu'\nu'\alpha'\beta'}=L^{\mu'}_{\ \mu}L^{\nu'}_{\ \nu}L^{\alpha'}_{\ \alpha}L^{\beta'}_{\ \beta}\,\chi^{\mu\nu\alpha\beta}\ ,
\end{equation}
and that the tensors $F_{\alpha\beta}$ and $W^{\alpha\beta}$ behave in a similar manner. For many applications, it is also useful to decompose $\dd \chi$ by means of the matrix identity
\begin{equation}\label{symmantisymm}
 \dd \chi  = \frac{\dd \chi+\ddind{\chi}{\hspace{2pt}T}}{2} + \frac{\dd \chi-\ddind{\chi}{\hspace{2pt}T}}{2}=\dd \chi_{\textindex{Symm.}}+{^{(2)}}\dd \chi\ ,       
\end{equation}
whereby one can isolate a symmetric contribution $\dd \chi_{\textindex{Symm.}}$ and an antisymmetric contribution ${^{(2)}}\dd \chi$\ \footnote{The reason for labeling the antisymmetric contribution $^{(2)}$ will become clear soon. Ultimately, we are aiming to match the notation of \cite{Hehl:2003}.}. A further split, reading
\begin{equation}\label{permutesplit}
 \dd \chi_{\textindex{Symm.}}  = {^{(1)}}\dd \chi+{^{(3)}}\dd \chi\ ,       
\end{equation}
can be achieved by requiring that ${^{(1)}}\chi^{[\mu\nu\alpha\beta]}$ vanishes, while ${^{(3)}}\chi^{[\mu\nu\alpha\beta]}$ does not\footnote{The square brackets used here denote index alternation \cite{Post}. One should also notice that the permutation ${^{(2)}}\chi^{[\mu\nu\alpha\beta]}$ (cf. Eq. \eqref{symmantisymm}) is identically zero.}. By virtue of \eqref{symmantisymm} and \eqref{permutesplit}, one can finally separate the medium response into a ``principal'' part, a ``skewon'' part and an ``axion'' part:
\begin{equation}
\label{chi123}
\chi^{\mu\nu\alpha\beta}={^{(1)}}\chi^{\mu\nu\alpha\beta}+{^{(2)}}\chi^{\mu\nu\alpha\beta}+{^{(3)}}\chi^{\mu\nu\alpha\beta}\ ,
\end{equation}
where ${^{(3)}}\chi^{\mu\nu\alpha\beta}=\alpha\epsilon^{\mu\nu\alpha\beta}$, and $\epsilon^{\mu\nu\alpha\beta}$ is the Levi-Civita tensor defined via
\begin{align}
  \epsilon^{\mu\nu\alpha\beta}&=g^{\mu\rho}g^{\nu\sigma}g^{\alpha\eta}g^{\beta\theta}\epsilon_{\rho\sigma\eta\theta}\ ,\\
  \epsilon_{0123}&=[-\det(\dd g)]^{\frac{1}{2}}\ .
\end{align}
A non-zero  ${}^{(2)}\dd \chi$ component is invariably excluded, since it implies that any of the following common symmetries is broken:
\begin{equation}
\dd \varepsilon=\ddind \varepsilon {\hspace{2pt}T}\ ,\quad\quad \dd \mu=\ddind \mu {\hspace{2pt}T}\ , \quad\quad\dd \alpha=-\ddind{\beta}{\hspace{2pt}T} .
\end{equation}
By contrast, a finite ${}^{(3)}\dd \chi$ is still quite rare, but has been observed in nature \cite{Lange:2001,Hehl:2008PLA} -- thus overturning a popular dogma (``Post's constraint'', see \cite{Post, Lakhtakia:1995}). The axion field also finds a practical application in the perfect electromagnetic conductor (PEMC) proposed in \cite{Lindell:2005Em}. Given these considerations, the present paper assumes that all skewon contributions vanish.

The easiest way to specify the properties of a material is via the 36 entries of $\dd \chi$. However, when a symbolic calculation must be carried out, it is often more convenient to reduce the constitutive relation to an abstract, more compact form. In engineering, this is mostly attained by simplifying $\dd \varepsilon$, $\dd \mu$, $\dd \alpha$ and $\dd \beta$ separately, according to their individual symmetry properties \cite{serdyukov}. This strategy, however, is manifestly \textit{not} relativistic, as it involves manipulating the four 3$\times$3 sub-matrices of \eqref{6x6chi} independently. Therefore, at least for the purpose of this work, one must choose a different approach, where $\chi^{\mu\nu\alpha\beta}$ is decomposed in terms of simpler spacetime quantities. For instance, vacuum is characterized by
\begin{equation}\label{vacmedium}
  \chi_{0}^{\mu\nu\alpha\beta}=(\mu_{0}/\varepsilon_{0})^{-\frac{1}{2}}\left(g^{\mu\alpha} g^{\nu\beta} - g^{\mu\beta} g^{\nu\alpha}\right)\ ,
\end{equation}
where $g^{\alpha\beta}$ is the inverse of the background metric $g_{\alpha\beta}$. Relaxing some restrictions leads to a first class of non-birefringent materials: the Q-media \cite{Lindell:2004Qm}. These are given by
\begin{equation}
\label{Qmedium}
 \chi_{\textindex{Q}}^{\mu\nu\alpha\beta} = s_Q\left(Q^{\mu\alpha} Q^{\nu\beta} - Q^{\mu\beta} Q^{\nu\alpha}\right)\ ,
\end{equation}
with $s_{Q}=\pm1$, and provide three supplementary sources of flexibility. 
\emph{Firstly,} the tensor $Q^{\alpha\beta}$ is not a metric, since although it is invertible, it is not necessarily symmetric. In the context of this paper, this property has little relevance, since a vanishing skewon implies that either $\dd Q=-\ddind {Q}{\hspace{2pt}T}$ or $\dd Q=\ddind{Q}{\hspace{2pt}T}$. Insomuch as the former case leads to an unspecified wave (co-)vector \cite{Bergamin:2010Sm}, only the latter case can be accepted. Thus, $\dd Q$ can henceforth be considered to be a ``constitutive'' metric, converging to $\dd g$ for the vacuum \eqref{vacmedium}. Accordingly, media of type \eqref{Qmedium} will be referred to as ``metric media''.
\emph{Secondly,} the constitutive signature need not be $(3,1)$, that is, ``Lorentzian''. This attribute will prove essential in deriving a compact representation for all non-birefringent materials (Section \ref{sec:3}). 
\emph{Lastly,} the impedance of the medium in Eq.\ \eqref{Qmedium} can be calculated to be
\begin{equation}\label{Qimpedance}
 \sqrt{\frac{\vert\det({Q}^{-1}_{\alpha\beta})\vert}{-\det(g_{\alpha\beta})}}\ ,
\end{equation}
and is not necessarily equal to $(\mu_{0}/\epsilon_{0})^{\frac{1}{2}}$. Moreover, the sign $s_{Q}$ allows one to obtain a negative index of refraction \cite{Bergamin:2010Nr}. 

A further class of materials, which naturally extends Q-media, has been proposed by Wall\'en and Lindell \cite{Lindell:DiffForms, Lindell:2004Gq}. They modify \eqref{Qmedium} by including two ``bivectors'' (antisymmetric tensors) $\mbf A$ and $\bar{\mbf A}$:
\begin{equation}
\label{genQchi}
 \chi_{\textindex{genQ}}^{\mu\nu\alpha\beta} = s_Q\left(Q^{\mu\alpha} Q^{\nu\beta} - Q^{\mu\beta} Q^{\nu\alpha}\right) + A^{\mu\nu} \bar{A}^{\alpha\beta}\ ,
\end{equation}
where, in this article, one must enforce $\bar{\mbf A} = s_{A}\mbf A$ with $s_{A}=\pm1$, since $^{(2)}\dd \chi$ is assumed to vanish.

Formulating the constitutive relation is the first step in characterizing the behaviour of light. The second milestone consists in solving Maxwell's equations -- in general or, as in the present work, in two specific circumstances. When studying the structure of exotic spacetimes, the propagation of sharp electromagnetic fronts is considered \cite{Hehl:2003}. By contrast, when modeling laboratory materials, some approximate plane-wave analysis is implemented (the light's front-velocity in a table-top medium is always exactly $c$ \cite{Milonni:2002, Milonni:2005}). Both of these scenarios correspond to the so-called ``geometrical optics'' limit, whence a dispersion (``Fresnel'') relation, linking the angular frequency $\omega$ to the spatial frequency $\bm{k}$, is derived. Using frame-independent notation, the 4-wave co-vector $\mbf{K}=(-\omega,\mbf{k})$ is found to obey
\begin{equation}
\label{fresnel}
\epsilon_{\alpha\beta\gamma\delta}\epsilon_{\eta\theta\kappa\lambda} \chi^{\alpha\beta\eta\theta}\chi^{\gamma\mu\nu\kappa}\chi^{\delta\rho\sigma\lambda}K_{\mu} K_\nu K_\rho K_\sigma = 0\, ,\hspace{-0.3cm}
\end{equation}
that is, a quartic Fresnel polynomial, cubic in the medium parameters \cite{Hehl:2003,Rubilar:2002Le,Obukhov:2000Rt, Lindell:2005We, Itin:2009Dr}. By using Eq.\ \eqref{chi123}, one can also prove that the effect of the axion on this formula is zero. Consequently, it impossible to specify ${^{(3)}}\dd \chi$ by analyzing light propagation in bulk. 

A material is non-birefringent if the Fresnel polynomial becomes bi-quadratic:
\begin{equation}
\label{fresnelNBR}
 (G^{\alpha\beta} K_\alpha K_\beta)^2 =0\ ,
\end{equation}
where $\dd G$ is an ``optical'' metric.  For an explicit derivation of the non-birefringence conditions we refer to the literature \cite{Lammerzahl:2004Rl,Itin:2005Nb} (cf.\ Section \ref{sec:3D}). The Fresnel equation for a Q-medium is trivial ($G^{\alpha\beta}\!\propto\!Q^{\alpha\beta}$). However, that of its extension \eqref{genQchi} is not \cite{Lindell:DiffForms}:
\begin{equation}
 \!\!\left[Q^{\mu\nu} K_\mu K_\nu\right]\!\left[\left(\dd Q - s_A \mbf A \cdot \ddind{Q}{-1}\cdot \mbf A\right)^{\alpha\beta} K_\alpha K_\beta\right] = 0\ .
\end{equation}
Birefringence, as induced by the bivector $\mbf{A}$, can be eliminated by requiring that
\begin{equation}\label{genQnbr}
 \mbf A \cdot \ddind{Q}{-1}\cdot \mbf A = a \dd Q\ ,
\end{equation}
where the proportionality constant $a$ may vanish, but $a \neq s_A$. The generalized Q-media that satisfy \eqref{genQnbr} cover only a range of non-birefringent materials. Nonetheless, they provide essential guidance for the generic case.
\section{Geometric representation of all non-birefringent media}
\label{sec:3}
\subsection{\hspace{-0.2cm}From Schuller's classification to a compact form}\label{sec:3A}
Discovering specific examples of non-birefringent media is relatively easy. However, finding a general representation is -- at a first sight -- difficult, given the complicated structure of the Fresnel equation \eqref{fresnel} . Remarkably, a classification scheme recently proposed by Schuller et al.\ \cite{Schuller:2010Ap}, brings this task to the realm of possibility (at least for a vanishing skewon field). The key breakthrough comes from an in-depth analysis of the Segre types \cite{HallSGR,StephaniExact} emerging from the modified tensor 
\begin{equation}\label{modifiedtensor}
\kappa_{\mu\nu}{}^{\rho\sigma} = \frac12 \epsilon_{\mu\nu\alpha\beta} \chi^{\alpha\beta\rho\sigma}\ .
\end{equation}
As an outcome, 23 equivalence classes are obtained, each equipped with a normal form (or ``typical representative''), whereby a complete categorization is readily achieved -- an arbitrary ``symmetric'' constitutive tensor can be linked uniquely to a typical representative by using a change of frame. Simultaneously, the non-birefringence conditions developed in \cite{Lammerzahl:2004Rl,Itin:2005Nb} are invariant under any basis transformation, as dictated by the principle of relativity. Therefore, an observer independent coverage of all non-birefringent media can be fully attained from the 23 specific matrices of Ref.\ \cite{Schuller:2010Ap} -- complying, in this sense, with the main aim of this section.  

A few technical comments are in order.  Schuller et al.\ opt to classify the lower-indexed ``area metric'' $G_{\mu\nu\rho\sigma}$, rather than the upper-indexed $\chi^{\mu\nu\rho\sigma}$ which is favored here. Regardless of this choice, the normal forms that they derive do apply directly to $\dd \chi$. In addition, the representatives in Ref.\ \cite{Schuller:2010Ap} are subject to the condition $\det \dd \chi = 1$. We choose to temporarily suspend this constraint, since the procedures below never preclude us from achieving a unit determinant. As a final remark, one should notice that Schuller et al.\ employ an alternative 6$\times$6 index ordering -- compare equation (2) in Ref.\ \cite{Schuller:2010Ap} with equation \eqref{munutoImap} of this manuscript.

With the help of computational software\footnote{The results of this paper were calculated using Mathematica\textsuperscript{\textregistered}.\label{mathematica}}, the required non-birefringent classification is derived straightforwardly. The immediate outcome is a set of matrices, which provide a valid solution, but not a transparent answer (see Appendices \ref{sec:appA} and \ref{sec:appB}). In spite of this difficulty, one necessary -- but not sufficient -- condition does emerge clearly. The structure
\begin{align}
 \chi_{\textindex{NBR}}^{\mu\nu\rho\sigma} =&\, s_Q\left(Q^{\mu\rho} Q^{\nu\sigma} - Q^{\mu\sigma} Q^{\nu\rho}\right)\nonumber\\ +&\, s_A A^{\mu\nu} A^{\rho\sigma} + s_{\tilde A} \tilde{A}^{\mu\nu} \tilde{A}^{\rho\sigma} + \alpha \epsilon^{\mu\nu\rho\sigma}\ ,\label{allNBRchi}
\end{align}
encompasses efficiently all skewonless non-birefringent materials, although some polarization-dependent solutions are yet to be ruled out. A complete refinement will soon be obtained. Nonetheless, some essential remarks should not be avoided. Equation \eqref{allNBRchi} is designed to conveniently encode the optical metric, solely by means of $\dd Q$. Moreover, our compact form naturally generalizes Eq. \eqref{genQchi}, at the small price of including an innocuous axion offset. It is also important to realize that the representation \eqref{allNBRchi} is invariant under some ``bivector space'' transformations comprising $\mbf A$ and $\tilde{\mbf A}$. If $s_A s_{\tilde A} = 1$, trigonometric rotations of the form
\begin{equation}
\label{realrotations}
 \begin{pmatrix}
  \mbf A' \\ \tilde{\mbf A}'
 \end{pmatrix} = \begin{pmatrix}
                  \cos \phi & -\sin \phi\\\sin\phi & \cos \phi
                 \end{pmatrix} \cdot \begin{pmatrix}
                                      \mbf A \\ \tilde{\mbf A}
                                     \end{pmatrix}
\end{equation}
map the medium tensor onto itself. If $s_A s_{\tilde A} = -1$, proper rotations have to be replaced by hyperbolic rotations
\begin{equation}
\label{imrotations}
 \begin{pmatrix}
  \mbf A' \\ \tilde{\mbf A}'
 \end{pmatrix} = \begin{pmatrix}
                  \cosh \varphi & -\sinh \varphi\\-\sinh\varphi & \cosh \varphi
                 \end{pmatrix} \cdot \begin{pmatrix}
                                      \mbf A \\ \tilde{\mbf A}
                                     \end{pmatrix}\ .
\end{equation}

\subsection{\hspace{-0.2cm}Necessary and sufficient conditions imply duality}\label{sec:3B}
At this point, the crucial idea is to study the Fresnel equation of $\chi_{\textindex{NBR}}^{\mu\nu\rho\sigma}$, so as to find a true, minimal, non-birefringent representation. From Eq.\ \eqref{fresnel}:
\begin{multline}
\label{allNBRfresnel}
  \det[ \dd Q] \Bigl((\dd Q - s_Q s_A \mbf A \cdot \ddind{Q}{-1} \cdot \mbf A)^{\mu\nu} K_\mu K_\nu \times\\ \times (\dd Q - s_Q s_{\tilde A} \tilde{\mbf A} \cdot \ddind{Q}{-1}\cdot \tilde{\mbf A})^{\rho\sigma} K_\rho K_\sigma \\ - s_A s_{\tilde A} \left[(\mbf A \cdot \ddind{Q}{-1}\cdot \tilde{\mbf A})^{\mu\nu} K_\mu K_\nu\right]^2 \Bigr) = 0\ .
\end{multline}
Thus, the medium is non-birefringent if and only if the following conditions are met:
\begin{gather}
\label{ABconstrI}
 \mbf A \cdot \ddind{Q}{-1}\cdot \mbf A \cdot  \ddind{Q}{-1} = a_1 \mathds 1\ , \\
 \label{ABconstrIA}
 \tilde{\mbf A} \cdot \ddind{Q}{-1}\cdot \tilde{\mbf A}\cdot  \ddind{Q}{-1} = a_2 \mathds 1\ ,\\
\label{ABconstrII}
 \mbf A \cdot \ddind{Q}{-1}\cdot \tilde{\mbf A}\cdot  \ddind{Q}{-1} +  \tilde{\mbf A} \cdot \ddind{Q}{-1}\cdot \mbf A \cdot  \ddind{Q}{-1} = 2 a_3 \mathds 1\ .
\end{gather}
Here, the scalars $a_1$, $a_2$ and $a_3$ must always comply with:
\begin{equation}
 (1-s_Q s_A a_1)(1-s_Q s_{\tilde A}a_2) - s_A s_{\tilde A} a_3 \neq 0\ .
\end{equation}
A solution to the above constraints is easily found by exploiting the bivector identity
\begin{equation}\label{bividentity}
 \frac{1}{2} A^{\mu \rho} \epsilon^Q_{\rho\nu\alpha\beta} A^{\alpha\beta} = \pm \frac{\sqrt{|\det \mbf A|}}{\sqrt{|\det \dd Q|}} \delta^\mu_\nu\ ,
\end{equation}
where $\epsilon^Q_{\mu\nu\alpha\beta}$ is a Levi-Civita-type tensor, with dual $\epsilon_{Q}^{\mu\nu\alpha\beta}$:
\begin{align}
  \epsilon^{Q}_{0123}&=\vert\det(\dd Q)\vert^{-\frac{1}{2}}\ ,\\
  \epsilon_{Q}^{\mu\nu\alpha\beta}\!&=Q^{\mu\rho}Q^{\nu\sigma}Q^{\alpha\eta}Q^{\beta\theta}\epsilon^{Q}_{\rho\sigma\eta\theta}\ ,\\
 \Rightarrow\quad\quad\quad(\epsilon_{Q})^{IJ}\!&=\mbox{sgn}[\det(\dd Q)](\epsilon_{Q}^{-1})^{IJ}\ .
\end{align}
Applying the relation \eqref{bividentity} in Eqs.\ \eqref{ABconstrI} and \eqref{ABconstrIA}, yields our main results; namely,
\begin{align}
\label{self-dual-a}
 A^{\mu\nu} &= \frac{s_X}{2} \epsilon_Q^{\mu\nu\alpha\beta} (\ddind{Q}{-1} \cdot \mbf A \cdot \ddind{Q}{-1})_{\alpha\beta}:=s_X({^{*}\!\mbf A})^{\mu\nu}\ ,\\
 \label{self-dual-a-tilde}
 \tilde A^{\mu\nu} &= \frac{s_Y}{2} \epsilon_Q^{\mu\nu\alpha\beta} (\ddind{Q}{-1} \cdot \tilde{\mbf A} \cdot \ddind{Q}{-1})_{\alpha\beta}:=s_Y({^{*}\!\tilde{\mbf A}})^{\mu\nu}\ ,
\end{align}
that is, $\mbf A$ and $\tilde{\mbf A}$ must be selfdual ($s_{X}, s_{Y} = 1$) or anti-selfdual ($s_{X}, s_{Y} = -1$), with:  
\begin{align}\label{dualtwice}
  \mbf{A}&={^{**}}\!\mbf{A}\ , & \tilde{\mbf{A}}&={^{**}}\!\tilde{\mbf{A}}\ .
\end{align}
The additional condition \eqref{ABconstrII} is solved automatically, provided $s_X = s_Y$.

Rotating the constraints \eqref{ABconstrI}--\eqref{ABconstrII} by means of Eq.\ \eqref{realrotations}, determines that $a_3=0$ can be achieved whenever $s_A s_{\tilde A} = 1$. By contrast, if $s_A s_{\tilde A} = -1$, the hyperbolic rotations \eqref{imrotations} allow one to attain $a_3=0$ only when
\begin{equation}
 \frac{a_1+a_2}{a_3} \notin [-2,+2]\ .
\end{equation}

\subsection{\hspace{-0.2cm}A summary with all explicit solutions to duality}\label{sec:3C}
In summary it is found that the optical metric of any non-birefringent medium obeys $G^{\mu\nu}\!\propto\!Q^{\mu\nu}$. All non-birefringent, skewonless materials are characterized by a Q-medium ``core'', which encodes the optical metric. Two (anti-)selfdual bivectors and an axion term can be included as corrections, but they do not affect the dispersion relation. Any further discussion needs to distinguish between the three possible signatures of $\dd Q$.
\vspace{-0.6cm}
\subsubsection{Signature (3,1)}
\vspace{-0.3cm}
When the signature of $\dd Q$ is Lorentzian (3,1), the following identity is true for any bivector $\mbf \psi$ \cite{Hehl:2003}:
\begin{align}
  \mbf {\psi}=-{^{**}}\mbf{\psi}\ .
\end{align}
Correspondingly, equation \eqref{dualtwice} cannot be satisfied, unless $\mbf A \equiv 0$ and $\tilde{\mbf A}\equiv 0$. Then, any non-birefringent material, with hyperbolic Fresnel equation, is simply a metric medium -- complemented by an eventual axion. Some observations in this sense can also be found in the group theory analysis of Ref.\ \cite{Schuller:2010Ap}.
\vspace{-0.6cm}
\subsubsection{Signatures (4,0) and (2,2)}
\vspace{-0.3cm}
The remaining two signatures support finite (anti-) selfdual bivector corrections. These can be written in terms of the vectors $\bm \tau$, $\mbf u$ and $\mbf v$, such that:
\begin{align}
\label{algsolutionI}
 A^{\mu\nu} &= 2 \tau^{[\mu} u^{\nu]} + s_X \epsilon_Q^{\mu\nu\rho\sigma} (\ddind{Q}{-1}\cdot \mbf{\tau})_{\rho}(\ddind{Q}{-1}\cdot \mbf u)_{\sigma}\ ,\\
 \label{algsolutionII}
 \tilde A^{\mu\nu} &= 2 \tau^{[\mu} v^{\nu]} + s_X \epsilon_Q^{\mu\nu\rho\sigma} (\ddind{Q}{-1}\cdot \mbf{\tau})_{\rho}(\ddind{Q}{-1}\cdot \mbf v)_{\sigma}\ .
\end{align}
In this parametrization, $\bm \tau$ singles out a particular direction, according to which all other vectors can be decomposed. In this sense, $\bm \tau$ plays a role similar to that of a time vector, and can be selected independently of $\mbf A$ and $\tilde{\mbf A}$ -- in almost all cases. Complications arise only if the signature is (2,2) and $\bm \tau$ is light-like, a scenario that we shall resolve later. Thus, one can take
\begin{equation}
\label{taulightlike}
 \bm \tau \cdot \ddind{Q}{-1} \cdot \bm \tau \neq 0\ ,
\end{equation}
while no specific signature is assumed yet. The components of $\mbf u$ and $\mbf v$ parallel to $\bm \tau$ never contribute to Eqs.\ \eqref{algsolutionI} and \eqref{algsolutionII}. Therefore, given that we excluded a light-like $\bm \tau$, one can choose $\bm \tau \cdot \ddind{Q}{-1} \cdot \mbf u = 0$ and $\bm \tau \cdot \ddind{Q}{-1} \cdot \mbf v = 0$, with no loss of generality. With these simplifications, it is possible to derive the following relations:
\begin{align}
 a_1 &= - (\bm \tau \cdot \ddind{Q}{-1} \cdot \bm \tau) (\mbf u \cdot \ddind{Q}{-1} \cdot \mbf u)\ , \\
 a_2 &= - (\bm \tau \cdot \ddind{Q}{-1} \cdot \bm \tau) (\mbf v \cdot \ddind{Q}{-1} \cdot \mbf v)\ , \\
 a_3 &= - (\bm \tau \cdot \ddind{Q}{-1} \cdot \bm \tau) (\mbf u \cdot \ddind{Q}{-1} \cdot \mbf v)\ .
\end{align}

If $\dd Q$ has signature (4,0), a true norm is established. Thus, the occurrence of $a_1=0$ or $a_2=0$ is forbidden, unless $\bm \tau$, $\mbf u$ or $\mbf v$ vanish. By means of trigonometric or hyperbolic rotations, the orthogonality $\mbf u \cdot \ddind{Q}{-1} \cdot \mbf v = 0$ can be obtained without altering $\dd \chi_{\textindex{NBR}}$. Equivalently, $a_3=0$ can always be assumed. Hence, for signature (4,0), the three vectors $\bm \tau$, $\mbf u$ and $\mbf v$ can be chosen to be mutually Q-orthogonal, with no effect on the medium response.

If $\dd Q$ has signature (2,2), the situation is more complicated. The vectors $\mbf u$ and $\mbf v$ can be light-like, thereby allowing $a_1=0$ and $a_2=0$. Moreover, if $s_A s_{\tilde A} = -1$, one can find cases where $a_3 = 0$ is no longer achievable.

Finally, we should comment on the possibility of a light-like $\bm \tau$. This choice is more subtle since there exist pairs of vectors $(\bm \tau,\mbf u)$ obeying all of the following conditions:
\begin{align}
 \tau^{[\mu} u^{\nu]} &\neq 0\ , & \bm \tau \cdot \ddind{Q}{-1} \cdot \mbf u &= 0\ , \\ \bm \tau \cdot \ddind{Q}{-1} \cdot \bm \tau &= 0\ , &\mbf u\cdot \ddind{Q}{-1} \cdot \mbf u &= 0\ .
\end{align}
Pairs of this kind span the kernel of the map \eqref{algsolutionI}, which links two vectors to one bivector. Whilst in specific situations one could allow for a light-like $\bm \tau$, this choice is inconvenient in general.
\subsection{\hspace{-0.2cm}A worked example: revision and further details}\label{sec:3D}
A worked example is proposed, which provides hands-on revision for Sections \ref{sec:3A}--\ref{sec:3C}. By implementing the prescriptions of Refs.\ \cite{Lammerzahl:2004Rl, Itin:2005Nb}, a specific ``Schuller'' normal form is explicitly reduced to the non-birefringent limit \eqref{allNBRchi}. Two bivectors, playing the role of $A^{\alpha\beta}$ and $\tilde{A}^{\alpha\beta}$, are readily determined. Their structure is shown to be encompassed by Eqs.\ \eqref{algsolutionI} and \eqref{algsolutionII}, thus illustrating a specific instance of (anti-)selfduality. 

As explained in Section \ref{sec:3A}, choosing a typical representative singles out a Segre type. In turn, this identifies a set of eigenvalues for the modified medium tensor of Eq.\ \eqref{modifiedtensor}; in this example, we select the characteristics
\begin{equation}\label{eigenvalueskappa}
  \sigma_{1}+\mbox{i}\tau_{1},\quad\sigma_{1}-\mbox{i}\tau_{1},\quad \lambda_{1},\quad\lambda_{2},\quad\lambda_{3},\quad\lambda_{4},
\end{equation}
where $\sigma_{1},\tau_{1},\lambda_{1},\dots,\lambda_{4}$ are real scalars. The associated 6$\times$6 matrix normal form $\chi_{\textindex{NF}}^{IJ}$ reads 
\begin{equation}
  \label{metaclassvi}
  \dd{\chi}_{\textindex{NF}}=\left[
\begin{array}{cccccc}
 -\tau _1 & 0 & 0 & \sigma _1 & 0 & 0 \\
 0 & \lambda _3 & 0 & 0 & \lambda _4 & 0 \\
 0 & 0 & \lambda _1 & 0 & 0 & \lambda _2 \\
 \sigma _1 & 0 & 0 & \tau _1 & 0 & 0 \\
 0 & \lambda _4 & 0 & 0 & \lambda _3 & 0 \\
 0 & 0 & \lambda _2 & 0 & 0 & \lambda _1
\end{array}
\right]\ ,
\end{equation}
as reported in Ref.\ \cite{Schuller:2010Ap} (cf.\ ``metaclass VI'' therein, with the index ordering \eqref{munutoImap}). At this point, the optical properties of the material are fully specified and the Fresnel equation \eqref{fresnel} can be invoked to obtain the polynomial
\begin{equation}
  \label{fresexpoly}
  M_{0}q^{4}+M_{1}q^{3}+M_{2}q^{2}+M_{3}q+M_{4}=0\ ,
\end{equation}
where $q$ is a component picked arbitrarily from the wave co-vector $K_{\mu}$. Crucially, the coefficients $M_{0},\dots,M_{4}$ are still a function of $(-\omega,\bm{k})$; however, they are independent of the particular entry $K_{\nu}$ such that $q\!=\!K_{\nu}$. For the material under consideration, if $q$ coincides with $\!K_{1}$:
\begin{align}
  M_0=&-\lambda _1 \lambda _3\tau _1\ ,\label{M0}\\
  M_1=&\ 0\ ,\label{M1}\\
  M_2=&+K_{0}^2 \tau _1\left[-\lambda _1^2+\left(\lambda _2-\lambda_4\right){}^2-\lambda _3^2\right]\nonumber\\
      &-K_{3}^2 \lambda _3 \left[-\lambda _1^2+\left(\lambda _2-\sigma_1\right){}^2+\tau _1^2\right]\nonumber\\
      &-K_{2}^2 \lambda _1 \left[-\lambda_3^2+\left(\lambda _4-\sigma _1\right){}^2+\tau _1^2\right]\ ,\label{M2}\\
  M_3=&-2 K_{0} K_{2} K_{3}\ \times\nonumber\\
\big\{\!&+\lambda _2^2 \left(\lambda _4-\sigma _1\right)-\lambda _1^2 \left(\lambda _4-\sigma _1\right)\nonumber\\
      &-\sigma _1 \left[\lambda _3^2-\lambda _4 \left(\lambda _4-\sigma _1\right)\right]\nonumber\\
      &-\lambda _4 \tau _1^2+\lambda _2 \left(\lambda _3^2-\lambda _4^2+\sigma _1^2+\tau _1^2\right)\!\big\}\ ,\label{M3}\\
  M_4=&-K_{0}^4 \lambda _1 \lambda _3 \tau _1\ +K_{2}^4 \lambda _1 \lambda _3 \tau _1\ +K_{3}^4 \lambda _1 \lambda _3 \tau _1\nonumber\\
      &+K_{2}^2 K_{3}^2\tau _1\left[\lambda _1^2+\lambda _3^2-\left(\lambda _2-\lambda _4\right){}^2\right]\nonumber\\
      &-K_{0}^2K_{2}^2 \lambda _3\left[\tau _1^2-\lambda _1^2+\left(\lambda _2-\sigma _1\right){}^2\right]\nonumber\\
      &-K_{0}^2K_{3}^2 \lambda _1 \left[\tau _1^2-\lambda _3^2+\left(\lambda _4-\sigma _1\right){}^2\right]\ ,\label{M4}
\end{align}
as one can verify by using a numerical software (see Footnote \ref{mathematica}). Two topics ought to be addressed: firstly, the identifications \eqref{M0}--\eqref{M4} depend on the choice of $q$, in contrast with the relativistic standard retained so far; secondly, birefringence must still be eliminated through the reduction of Eq.\ \eqref{fresexpoly} to the form \eqref{fresnelNBR}. 

\begin{figure}
 \centering
    \includegraphics[width=\columnwidth]{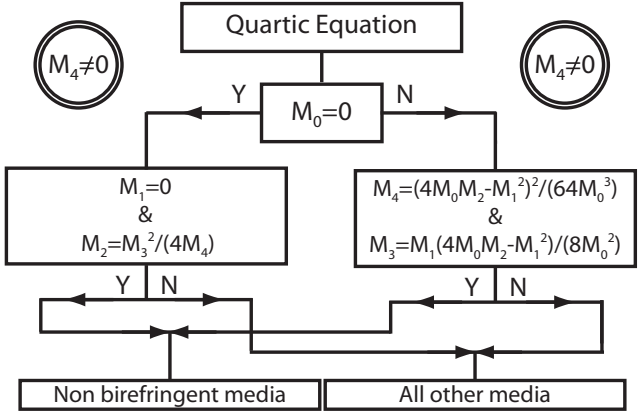}  
\caption{\label{NBRscheme} The procedure to make Eq.\ \eqref{fresexpoly} a bi-quadratic. One must check that $M_{4}\!\neq\!0$ at every node, so as to avoid $q\!=\!0$. The birefringence elimination scheme proposed in Refs.\ \cite{Lammerzahl:2004Rl, Itin:2005Nb} does not include the $M_{0}\!=\!0$ branch depicted on the left.}
\end{figure}

Remarkably, once a bi-quadratic $(G^{\alpha\beta} K_\alpha K_\beta)^{2}\!=\!0$ is derived, the result is independent of how $q$ is defined. This fact can be verified directly, by means of a computer, or geometrically from the uniqueness of the ``light-cone'' \cite{Lammerzahl:2004Rl}. Knowing that (when birefringence is ruled out) a frame independent outcome is ensured, one can safely apply the scheme of Figure \ref{NBRscheme} and retrieve Eq.\ \eqref{fresnelNBR}.    

The analysis underpinning this method is due to L\"ammerzahl, Hehl \cite{Lammerzahl:2004Rl} and Itin \cite{Itin:2005Nb}. The present example demonstrates that, in extension to the existing theory, non-birefringent solutions can be obtained even for $M_{0}\!=\!0$ (Fig.\ \ref{NBRscheme}, left-hand branch). By contrast, $M_{4}$ must always be nonzero, otherwise the dispersion relation \eqref{fresexpoly} leads to the unphysical root $q\!=\!0$. Carrying out the procedure described in Figure \ref{NBRscheme} for the medium \eqref{metaclassvi} yields
\begin{align}
  M_{0}&=0\,, &\qquad\lambda_{1}&=0\,, &\qquad\lambda_{2}&=\pm\lambda_{3}+\lambda_{4}\,,\label{lambdaone}\\
M_{0}&=0\,, &\qquad\lambda_{3}&=0\,, &\qquad\lambda_{4}&=\pm\lambda_{1}+\lambda_{2}\,,\label{lambdathree}\\
M_{0}&=0\,, &\qquad\tau_{1}&=0\,, &\qquad\sigma_{1}&=\pm\lambda_{3}+\lambda_{4}\,,\label{taua}\\
M_{0}&=0\,, &\qquad\tau_{1}&=0\,, &\qquad\sigma_{1}&=\pm\lambda_{1}+\lambda_{2}\,,\label{taub}
\end{align}
whilst taking $M_{0}\!\neq\!0$ provides no results compatible with
\begin{equation}
M_4\neq\frac{\left(4M_0M_2-M_1^2\right)^{2}}{64M_0^{3}}\ . 
\end{equation}
It is worth noticing that the eigenvalues in Eq.\ \eqref{eigenvalueskappa} can be re-ordered, for example by exchanging $\lambda_{3}$ with $\lambda_{1}$ and $\lambda_{4}$ with $\lambda_{2}$. Accordingly, one can observe that Eqs.\ \eqref{lambdathree} and \eqref{taub} are equivalent to Eqs.\ \eqref{lambdaone} and \eqref{taua}, respectively. With no loss of generality, it is thus possible to focus on the two cases
\begin{align}
 \mbox{(i)}\qquad\quad\lambda_{3}&=0\,, &\ \lambda_{4}&=-\zeta\lambda_{1}+\lambda_{2}\,,\label{nbri}\\
 \mbox{(ii)}\qquad\quad\tau_{1}&=0\,, &\ \sigma_{1}&=+\zeta\lambda_{1}+\lambda_{2}\,,\label{nbrii}
\end{align}
with $\zeta\!=\!\pm1$. The corresponding Fresnel equations \eqref{fresnel} are non-birefringent, as required, and read
\begin{align}
-\lambda_{1}[(\lambda_{2}-\sigma_{1}-\zeta \lambda_{1})^{2}+\tau_{1}^{2}](K_{1}K_{2}\!+\!\zeta K_{0}K_{3})^{2}\!=&0,\\
-\lambda_{1}[(\lambda_{2}-\lambda_{4}+\zeta \lambda_{1})^{2}-\lambda_{3}^{2}](K_{1}K_{2}\!+\!\zeta K_{0}K_{3})^{2}\!=&0.
\end{align}
Moreover, the pre-factors multiplying $(K_{1}K_{2}\!+\!\zeta K_{0}K_{3})^{2}$ must be non-zero, otherwise the dispersion relation vanishes identically. One thus enforces the conditions
\begin{align}
  \mbox{(i)}\,\qquad\lambda_{1}&\neq0, & (\lambda_{2}-\sigma_{1}-\zeta \lambda_{1})^{2}+\tau_{1}^{2}&\neq0, \label{constraini}\\
\mbox{(ii)}\qquad\lambda_{1}&\neq0, & (\lambda_{2}-\lambda_{4}+\zeta \lambda_{1})^{2}-\lambda_{3}^{2}&\neq0,\label{constrainii}
\end{align}
so as to identify the optical metrics for cases $\mbox{(i)}$ and $\mbox{(ii)}$:
\begin{equation}
  \dd{G}_{\textindex{(i)}}=\dd{G}_{\textindex{(ii)}}=\left(
    \begin{array}{cccc}
      0 & 0 & 0 & 1\\
      0 & 0 & \zeta & 0\\
      0 & \zeta & 0 & 0\\
      1 & 0 & 0 & 0
    \end{array}
\right)\ .\label{gigii}
\end{equation}
As an aside, one should observe that $\dd{G}_{\textindex{(i)}}$ and $\dd{G}_{\textindex{(ii)}}$ have signature $(2,2)$. Hence, these metrics are not suitable for governing the causal structure of spacetime \cite{Schuller:2010Ap, Ratzel:2010}. All realizations of Eq.\ \eqref{gigii} are restricted to laboratory materials, in the steady-state limit (Section \ref{sec:2}). In agreement with the findings of Section \ref{sec:3A}, the non-birefringent media \eqref{nbri} and \eqref{nbrii} have a constitutive relation of the form \eqref{allNBRchi}. More specifically, one can show that   
\begin{align}
  \chi_{\textindex{(i)}}^{\mu\nu\rho\sigma} =&-\mbox{sgn}(\lambda_{1})\left(Q_{\textindex{(i)}}^{\mu\rho} Q_{\textindex{(i)}}^{\nu\sigma} - Q_{\textindex{(i)}}^{\mu\sigma} Q_{\textindex{(i)}}^{\nu\rho}\right)\nonumber\\ &+s_{\textindex{(i)}} A_{\textindex{(i)}}^{\mu\nu} A_{\textindex{(i)}}^{\rho\sigma}\: + \tilde{s}_{\textindex{(i)}} \tilde{A}_{\textindex{(i)}}^{\mu\nu} \tilde{A}_{\textindex{(i)}}^{\rho\sigma}\: + \lambda_{2} \epsilon^{\mu\nu\rho\sigma}\ ,\\
 \chi_{\textindex{(ii)}}^{\mu\nu\rho\sigma} =&-\mbox{sgn}(\lambda_{1})\left(Q_{\textindex{(ii)}}^{\mu\rho} Q_{\textindex{(ii)}}^{\nu\sigma} - Q_{\textindex{(ii)}}^{\mu\sigma} Q_{\textindex{(ii)}}^{\nu\rho}\right)\nonumber\\ &+s_{\textindex{(ii)}} A_{\textindex{(ii)}}^{\mu\nu} A_{\textindex{(ii)}}^{\rho\sigma} + \tilde{s}_{\textindex{(ii)}} \tilde{A}_{\textindex{(ii)}}^{\mu\nu} \tilde{A}_{\textindex{(ii)}}^{\rho\sigma} + \lambda_{2} \epsilon^{\mu\nu\rho\sigma}\ ,
\end{align}
where the two constitutive metrics $Q_{\textindex{(i)}}^{\alpha\beta}$ and $Q_{\textindex{(ii)}}^{\alpha\beta}$ coincide  
\begin{align}
  \dd{Q}_{\textindex{(i)}}&=\sqrt{\vert\lambda_1\vert}\ \dd{G}_{\textindex{(i)}}, & \dd{Q}_{\textindex{(ii)}}&=\sqrt{\vert\lambda_1\vert}\ \dd{G}_{\textindex{(ii)}},
\end{align}
and are non-singular by virtue of Eqs.\ \eqref{constraini} and \eqref{constrainii}. The bivector corrections $\bm{A}_{\textindex{(i)}},\tilde{\bm{A}}_{\textindex{(i)}},\bm{A}_{\textindex{(ii)}}$ and $\tilde{\bm{A}}_{\textindex{(ii)}}$ are given by
\begin{align}
  \bm{A}_{\textindex{(i)}}\!=&\!\left[
\begin{array}{cccc}
 0 & a_{\textindex{(i)}} & 0 & 0 \\
 -a_{\textindex{(i)}} & 0 & 0 & 0 \\
 0 & 0 & 0 & b_{\textindex{(i)}} \\
 0 & 0 & -b_{\textindex{(i)}} & 0
\end{array}
\right]\ , \label{firstbivsix}\\
\tilde{\bm{A}}_{\textindex{(i)}}\!=&\!\left[
\begin{array}{cccc}
 0 & c_{\textindex{(i)}} & 0 & 0 \\
 -c_{\textindex{(i)}} & 0 & 0 & 0 \\
 0 & 0 & 0 & d_{\textindex{(i)}} \\
 0 & 0 & -d_{\textindex{(i)}} & 0
\end{array}
\right]\ , \\ 
\bm{A}_{\textindex{(ii)}}\!=&\!\left[
\begin{array}{cccc}
 0 & 0 & a_{\textindex{(ii)}} & 0 \\
 0 & 0 & 0 & b_{\textindex{(ii)}} \\
 -a_{\textindex{(ii)}} & 0 & 0 & 0 \\
 0 & -b_{\textindex{(ii)}} & 0 & 0
\end{array}
\right]\ , \\
\tilde{\bm{A}}_{\textindex{(ii)}}\!=&\!\left[
\begin{array}{cccc}
 0 & 0 & c_{\textindex{(ii)}} & 0 \\
 0 & 0 & 0 & d_{\textindex{(ii)}} \\
 -c_{\textindex{(ii)}} & 0 & 0 & 0 \\
 0 & -d_{\textindex{(ii)}} & 0 & 0\label{lastbivsix}
\end{array}
\right]\ ,
\end{align}
where the real scalars $\{a_{\textindex{(i)}},\dots,d_{\textindex{(i)}}\}$ and $\{a_{\textindex{(ii)}},\dots,d_{\textindex{(ii)}}\}$ must be expressed in terms of the eigenvalues \eqref{eigenvalueskappa}. Similarly, the signs $\{s_{\textindex{(i)}},\tilde{s}_{\textindex{(i)}},s_{\textindex{(ii)}},\tilde{s}_{\textindex{(ii)}}\}$ are still to be specified. One can directly verify that, defining the variables
\begin{align}
  \rho_{\textindex{(i)\,}}&:=-\zeta\lambda_{1}-\lambda_{2}+\sigma_{1}\ ,\\
  \rho_{\textindex{(ii)}}&:=+\zeta\lambda_{1}-\lambda_{2}+\lambda_{4}\ ,\\
  \xi_{\textindex{(ii)}}&:=\mbox{sgn}(\vert\lambda_{3}\vert\!-\!\vert\rho_{\textindex{(ii)}}\vert)\ ,
\end{align}
and selecting the signs $s_{\textindex{(i)}},\tilde{s}_{\textindex{(i)}},s_{\textindex{(ii)}}$ and $\tilde{s}_{\textindex{(ii)}}$ according to  
\begin{align}
 s_{\textindex{(i)}\,}&=\mbox{sgn}(\tau_{1})\,, &  \tilde{s}_{\textindex{(i)}\,}&=-\:\:\mbox{sgn}(\tau_{1})\,\,, \\
s_{\textindex{(ii)}}&=\mbox{sgn}(\lambda_{3})\,, &  \tilde{s}_{\textindex{(ii)}}&=\xi_{\textindex{(i)}}\mbox{sgn}(\lambda_{3})\,, 
\end{align}
is compatible with the following choice of bivector entries:
\begin{align}
  a_{\textindex{(i)}}&=-\,d_{\textindex{(i)}}\ ,\\
  b_{\textindex{(i)}}&=+\,c_{\textindex{(i)}}\ ,\\
c_{\textindex{(i)}}&=+\,\mbox{sgn}(\tau_1)\ [(\vert\rho_{\textindex{(i)}}\vert^{2}+\vert\tau_{1}\vert^{2})^{\frac{1}{2}}+\vert\tau_{1}\vert]^{\frac{1}{2}}/\sqrt{2}\ ,\\
d_{\textindex{(i)}}&=-\,\mbox{sgn}(\rho_{\textindex{(i)}})[(\vert\rho_{\textindex{(i)}}\vert^{2}+\vert\tau_{1}\vert^{2})^{\frac{1}{2}}-\vert\tau_{1}\vert]^{\frac{1}{2}}/\sqrt{2}\ ,\\
a_{\textindex{(ii)}}&=+\,\xi_{\textindex{(ii)}}\,\mbox{sgn}(\lambda_3)\,\mbox{sgn}(\rho_{\textindex{(ii)}})\,b_{\textindex{(ii)}}\ ,\\
c_{\textindex{(ii)}}&=-\,\xi_{\textindex{(ii)}}\,\mbox{sgn}(\lambda_3)\,\mbox{sgn}(\rho_{\textindex{(ii)}})\,d_{\textindex{(ii)}}\ ,\\
b_{\textindex{(ii)}}&=+\left[\vert\lambda_{3}\vert-\xi_{\textindex{(ii)}}\vert\rho_{\textindex{(ii)}}\vert\right]^{\frac{1}{2}}/\sqrt{2}\ ,\\
d_{\textindex{(ii)}}&=+\left[\vert\rho_{\textindex{(ii)}}\vert+\xi_{\textindex{(ii)}}\vert\lambda_{3}\vert\right]^{\frac{1}{2}}/\sqrt{2}\ .
\end{align}
As a final consistency check, it is possible to verify that the conditions \eqref{self-dual-a} and \eqref{self-dual-a-tilde} are satisfied. In fact, one can demonstrate that
\begin{align}
 A^{\mu\nu}_{\textindex{(i)\phantom{i}}}=&2\tau_{\textindex{(i)\phantom{i}}}^{[\mu} u_{\textindex{(i)\phantom{i}}}^{\nu]}+\zeta \epsilon_{Q_{\textindex{(i)}}}^{\mu\nu\rho\sigma} (\ddind{Q}{-1}_{\textindex{(i)}}\cdot\bm{\tau}_{\textindex{(i)}})_{\rho}(\ddind{Q}{-1}_{\textindex{(i)}}\cdot\bm{u}_{\textindex{(i)}})_{\sigma}\,,\\
 \tilde A_{\textindex{(i)\phantom{i}}}^{\mu\nu}=&2\tau_{\textindex{(i)\phantom{i}}}^{[\mu}v_{\textindex{(i)\phantom{i}\,}}^{\nu]}+\zeta\epsilon_{Q_{\textindex{(i)}}}^{\mu\nu\rho\sigma} (\ddind{Q}{-1}_{\textindex{(i)}}\cdot\bm{\tau}_{\textindex{(i)}})_{\rho}(\ddind{Q}{-1}_{\textindex{(i)}}\cdot\bm{v}_{\textindex{(i)}})_{\sigma}\,,\\
A^{\mu\nu}_{\textindex{(ii)}}=&2\tau_{\textindex{(ii)}}^{[\mu}u_{\textindex{(ii)}}^{\nu]}+\zeta \epsilon_{Q_{\textindex{(ii)}}}^{\mu\nu\rho\sigma} (\ddind{Q}{-1}_{\textindex{(ii)}}\cdot\bm{\tau}_{\textindex{(ii)}})_{\rho}(\ddind{Q}{-1}_{\textindex{(ii)}}\cdot\bm{u}_{\textindex{(ii)}})_{\sigma}\,,\\
 \tilde A_{\textindex{(ii)}}^{\mu\nu}=&2\tau_{\textindex{(ii)}}^{[\mu}v_{\textindex{(ii)\,}}^{\nu]}+\zeta\epsilon_{Q_{\textindex{(ii)}}}^{\mu\nu\rho\sigma} (\ddind{Q}{-1}_{\textindex{(ii)}}\cdot\bm{\tau}_{\textindex{(ii)}})_{\rho}(\ddind{Q}{-1}_{\textindex{(ii)}}\cdot\bm{v}_{\textindex{(ii)}})_{\sigma}\,,
\end{align}
where the triplets $\{\bm{\tau}_{\textindex{(i)}},\bm{u}_{\textindex{(i)}},\bm{v}_{\textindex{(i)}}\}$ and $\{\bm{\tau}_{\textindex{(ii)}},\bm{u}_{\textindex{(ii)}},\bm{v}_{\textindex{(ii)}}\}$ are specified (non-uniquely) as 
\begin{align}
\bm{\tau}_{\textindex{(i)}\phantom{i}}=&\ (0,1,0,0)\ ,\\
\bm{u}_{\textindex{(i)\phantom{i}}}=&\ (a_{\textindex{(i)}}/15,0,0,b_{\textindex{(i)}}/17)\ ,\\
\bm{v}_{\textindex{(i)\phantom{i}}}=&\ (c_{\textindex{(i)}}/15,0,0,d_{\textindex{(i)}}/17)\ ,\\
\bm{\tau}_{\textindex{(ii)}}=&\ (0,0,1,0)\ ,\\
\bm{u}_{\textindex{(ii)}}=&\ (a_{\textindex{(ii)}}/17,0,0,b_{\textindex{(ii)}}/15)\ ,\\
\bm{v}_{\textindex{(ii)}}=&\ (c_{\textindex{(ii)}}/17,0,0,d_{\textindex{(ii)}}/15)\ .
\end{align}
Clearly, the bivectors \eqref{firstbivsix}--\eqref{lastbivsix} display the typical structure \eqref{algsolutionI}-\eqref{algsolutionII}. Thus, the corrections $\bm{A}_{\textindex{(i)}},\dots,\tilde{\bm{A}}_{\textindex{(ii)}}$ are selfdual for $\zeta=+1$, and anti-selfdual for $\zeta=-1$.  
\section{Unique light-cone and transformation optics}
\label{sec:4}
\subsection{\hspace{-0.2cm}The light-cone and premetric electrodynamics}
\label{sec:4A}
This section is organized as follows: firstly, some refinements are introduced, as required by the premetric electrodynamics of spacetime. Then, it is demonstrated that the results of Section \ref{sec:3} remain valid, besides all amendments. Finally, the work of Ref.\ \cite{Hehl:2003} is linked to that of Ref.\ \cite{Lammerzahl:2004Rl}.

Following the premetric formalism of Post \cite{Post}, the mathematics used so far needs one adjustment. The permutation symbols $e^{\alpha\beta\gamma\delta}$ and $\hat{e}_{\alpha\beta\gamma\delta}$, with
 \begin{align}
e^{0123}&=1\ , & \hat{e}_{0123}&=1\ ,
 \end{align}
\emph{must} be employed. Correspondingly, it is necessary to go beyond the rules of tensor transformation, since
\begin{equation}\label{densityperm}
  e^{\alpha'\beta'\gamma'\delta'}=\vert\det(L^{\rho'}_{\ \rho})\vert^{-1}L^{\alpha'}_{\ \alpha}L^{\beta'}_{\ \beta}L^{\gamma'}_{\ \gamma}L^{\delta'}_{\ \delta}\, e^{\alpha\beta\gamma\delta}\ ,
\end{equation}
where the ``density'' factor $\vert\det(L^{\rho'}_{\ \rho})\vert^{-w}$ is said to have ``weight'' $w\!=\!+1$. Given that the premetric medium response $\dd \chi_{\textindex{PM}}$ is a derived quantity \cite{Hehl:2003}, 
\begin{equation}\label{chipm}
  \chi_{\textindex{PM}}^{\mu\nu\alpha\beta}:=\frac{1}{2}e^{\mu\nu\rho\sigma}\kappa_{\rho\sigma}{}^{\alpha\beta}\ ,
\end{equation}
the following ``amended'' transformation rule is obtained (compare with Eq.\ \eqref{chiframetransform})
\begin{equation}\label{chipmtransf}
   \chi_{\textindex{PM}}^{\mu'\nu'\alpha'\beta'}=\vert\det(L^{\rho'}_{\ \rho})\vert^{-1}L^{\mu'}_{\ \mu}L^{\nu'}_{\ \nu}L^{\alpha'}_{\ \alpha}L^{\beta'}_{\ \beta}\,\chi_{\textindex{PM}}^{\mu\nu\alpha\beta}\ ,
\end{equation}
so that $\chi_{\textindex{PM}}^{\mu\nu\alpha\beta}$ is a tensor density with weight $w=+1$. 

At this point, we verify that the refinements \eqref{densityperm} and \eqref{chipmtransf} do not affect the findings of Section \ref{sec:3}. Crucially, one can prove (cf.\ Appendix \ref{sec:appC}) that Schuller's classification still applies, and that the representation \eqref{allNBRchi} is translated to:
\begin{align}\label{allNBRchiPM}
 \chi_{\textindex{PM}}^{\mu\nu\rho\sigma} =&\sqrt{\vert\det(Q^{-1}_{\alpha\beta})\vert}M\bigl[\left(Q^{\mu\rho} Q^{\nu\sigma} - Q^{\mu\sigma} Q^{\nu\rho}\right)\nonumber\\ +&\, s_A A^{\mu\nu} A^{\rho\sigma} + s_{\tilde A} \tilde{A}^{\mu\nu} \tilde{A}^{\rho\sigma}\bigr] + \alpha e^{\mu\nu\rho\sigma}\ ,
\end{align}
where $M$ is a true scalar. The ``differences'' between Eq.\ \eqref{allNBRchiPM} and the counterpart \eqref{allNBRchi} drop out from the Fresnel equation. Hence, the expression \eqref{allNBRfresnel} is always obtained, together with the conditions \eqref{self-dual-a} and \eqref{self-dual-a-tilde}. Consequently, all the conclusions of Section \ref{sec:3} remain valid.

To understand the importance of our work in the context of premetric electrodynamics, a few results of this program should be mentioned. Hehl and Obukhov \cite{Hehl:2003} consider some key experimental facts, and thus deduce a constitutive metric -- a ``light-cone'' structure -- emerging from the response $\dd \chi_{\textindex{PM}}$. By enforcing electric-magnetic reciprocity, they require the axion free contribution $\tilde{\chi}^{IJ}_{\textindex{PM}}$ to obey the closure condition
\begin{equation}\label{closurecondition}
\hat{e}_{IK} \tilde \chi_{\textindex{PM}}^{KL} \hat{e}_{LM} \tilde \chi_{\textindex{PM}}^{MJ} = \tilde \kappa_{I}{}^{K} \tilde \kappa_{K}{}^{J} = -\left(1/\tilde{Z}^{2}\right)\delta_I^J\ , 
\end{equation}
where $\tilde{Z}$ is a constant. As a second and final constraint, they also set the skewon part to zero, so as to obtain the desired Lorentzian metric -- together with an arbitrary axion contribution. In Ref.\ \cite{Lammerzahl:2004Rl} L\"ammerzahl and Hehl exploit different observational facts and develop a parallel, yet unrelated, derivation. Rather than simplifying the material response directly, they study the characteristics of the Fresnel equation, thus achieving an optical metric. They impose that:
\begin{itemize}
 \item  The medium is non-birefringent, that is, there exists a factorization of the form \eqref{fresnelNBR}. Hence, an optical metric is uniquely defined. 
 \item Given a value of $\omega$, the optical metric supports two real (``running-wave'') solutions -- accordingly, the Fresnel polynomial is hyperbolic.
\end{itemize}
Clearly, the two schemes presented above are conceptually distant. Ref.\ \cite{Hehl:2003} proceeds by exploiting symmetries, and deduces a constitutive metric $\dd Q$. By contrast, Ref.\ \cite{Lammerzahl:2004Rl} takes a strictly ``eikonal'' route, and derives an optical metric $\dd G$. Given these differences, it is remarkable that, in the absence of skewon, we can indeed establish a link: as seen in Section \ref{sec:3}, the optical metric for all non-birefringent media is proportional to the constitutive metric, $Q^{\alpha\beta}\propto G^{\alpha\beta}$. Observing that vacuum must have signature $(3,1)$\,\footnote{In empty space, causality demands the Fresnel equation to be strictly hyperbolic \cite{Schuller:2010Ap, Ratzel:2010}. For table-top materials, this restriction can be relaxed, since we assumed continuous wave operation -- rather than information transport, as encoded in the front velocity (Section \ref{sec:2}).}, the result $\mbf A\equiv\tilde{\mbf A}\equiv0$ is found immediately. Thus, the distinction between constitutive and optical structures disappears:
\begin{equation}
\!\!\chi_{\textindex{PM}}^{\mu\nu\rho\sigma}\!=\frac{\tilde{Z}^{-1}}{\sqrt{-\det G^{\alpha\beta}}}\left[G^{\mu\rho} G^{\nu\sigma}\!\!-\!G^{\mu\sigma} G^{\nu\rho}\right]+\alpha e^{\mu\nu\rho\sigma},\label{ifclosurethen}
\end{equation}
up to a choice of $\tilde{Z}$ and $\alpha$. These residual sources of flexibility do not distinguish the works of Ref.\ \cite{Hehl:2003} and Ref.\ \cite{Lammerzahl:2004Rl}. Indeed, Hehl and Obukhov ``manually'' set $\tilde{Z}=(\mu_{0}/\epsilon_{0})^{1/2}$ and $\alpha=0$ in (D.6.11) and (D.6.12).

\subsection{\hspace{-0.2cm}Non-birefringent transformation optics}
\label{sec:4B}
Transformation optics (TO) is a useful design tool, which offers direct control over the light rays \cite{Pendry:2006Sc,Leonhardt:2006Nj}. Its working principle is simple and can be explained in few sentences. The starting point is always a well-understood setup, called ``base-geometry",  supporting a pre-determined electromagnetic configuration. A coordinate change, or ``deformation", is then applied, which drags the fields to a useful layout -- and which specifies the necessary medium parameters. Thereby, a powerful recipe is spelled out, which can be used to engineer materials with tailored optical properties. Novel devices can thus be developed in the laboratory, whilst dissolving any reference to the enabling grid-distortion.  

Most applications of TO, such as cloaking \cite{Pendry:2006Sc, Leonhardt:2006Sc, Li:2008Cc}, light-harvesting \cite{Aubry:2010Nl} and lensing \cite{Pendry:2003Cl}, make use of a simple base-geometry: usually vacuum, or an arrangement of scalar $\varepsilon$ and $\mu$.   By virtue of this choice, birefringence is always ruled out from the very outset and never affects the design process. For many purposes, this is a true benefit, which should be exploited fully;  whence the question:  what is the most general substrate for constructing a TO scheme free of birefringence? 

In the case of zero skewon, a complete answer comes from the analysis of Section \ref{sec:3}. In particular, any material featured in the base-geometry must take the compact form \eqref{allNBRchi}, whilst obeying the duality rules \eqref{self-dual-a} and \eqref{self-dual-a-tilde}. A number of interesting scenarios are encompassed in this general solution -- for example, when the signature is (4,0) or (2,2). Yet it appears that, technologically, the Lorentzian metric does enjoy a special status, since it guarantees propagating bulk modes.  Hence, under most circumstances, the Fresnel equation is hyperbolic, and the bivectors $\mbf{A}$ and $\tilde{\mbf{A}}$ vanish. Accordingly, non-birefringent TO is reduced to the simple class of (3,1)-metric media \eqref{Qmedium}, complemented by an axion. 

Given this last consideration, it is easy to understand why avoiding birefringence requires the TO substrate to be vacuum, with only few exceptions. A comparison with $\dd\chi_{0}$ in Eq.\ \eqref{vacmedium} reveals that, at this point, only two quantities can still be adjusted: the impedance \eqref{Qimpedance}, with the sign $s_Q$, and an axion term. Even so, when probing an interface with empty space, selecting
\begin{align}
s_{Q}\sqrt{\frac{\det({Q}^{-1}_{\alpha\beta})}{\det(g_{\alpha\beta})}}&\neq\sqrt{\frac{\mu_{0}}{\epsilon_{0}}}\ ,
\end{align}
or choosing $^{(3)}\dd\chi\neq0$, generates finite reflections in trivial scenarios (see, for example, Ref.\ \cite{Obukhov:2005Pl}). By observing that TO does not naturally account for back-scattered waves, one can conclude that vacuum is, indeed, the required base-geometry.

As a final remark, it is also important to notice that both the impedance and the axion may be chosen to be spacetime dependent -- a case that is not easily covered by TO, or geometrical optics. 

\section{Conclusions}
\label{sec:5}
Starting from the classification of all skewonless, linear media by Schuller et al.\ \cite{Schuller:2010Ap}, we derived an intuitive representation for all non-birefringent materials: \emph{a metric medium core was employed, to which three ``dispersion-relation preserving'' corrections were added}. 

This result provides a powerful inverse design tool. Once the optical metric of a material is known from propagation, the constitutive metric is fixed up to a constant: enforcing ${Q}^{\alpha\beta}\!\propto\!{G}^{\alpha\beta}$ is a \emph{very} important step. Then, some freedom is still available in the corrections, namely an axion and two (anti-)selfdual bivectors ($\mbf{A}$ and $\tilde{\mbf A}$).

An explicit form for $\mbf{A}$ and $\tilde{\mbf A}$ was also developed, and was found to depend on the metric's signature. Remarkably, the (anti-)selfdual bivectors were observed to vanish in the Lorentzian (3,1) case. Accordingly, a reduction of all hyperbolic, skewonless, non-birefringent materials to metric media was achieved. The impact of this result both in premetric electrodynamics and in transformation optics was demonstrated (Section \ref{sec:4}).

Considering further research, it would be interesting to study the medium tensor \eqref{allNBRchi} more in detail. Indeed, the Fresnel equation \eqref{allNBRfresnel} features interesting examples of separable (uniaxial) and non separable birefringent media. In this context, one might even wonder whether the structure \eqref{allNBRchi} covers all skewonless media. On a different note, the inclusion of a finite skewon in our analysis would also constitute a stimulating challenge. At the level of the Fresnel equation, a skewon term can be accounted for straightforwardly, thanks to the separation (D.2.42) in Ref. \cite{Hehl:2003}. Additionally, a Segre classification of the full medium tensor still exists, even if $\chi^{IJ}$ cannot be diagonalized in general. Nonetheless, it is questionable that the ensuing classification remains manageable in a simple way.

We believe that this work could spark some interest beyond the linear-optics community. For instance, Eq.\ \eqref{allNBRchi} does encompass the nonlinear medium of Obukhov and Rubilar \cite{Obukhov:2002Prd} (in the limit where sharp fronts propagate with no birefringence). One can confirm this statement by observing that the material in Sec.\ 8 of Ref.\ \cite{Obukhov:2002Prd} is skewonless and obeys the closure relation \eqref{closurecondition}. Accordingly, the simple optical response \eqref{ifclosurethen} is retrieved, as explained in the manual \cite{Hehl:2003}. One should finally notice that our findings could be relevant to the study of QED-induced spacetime structures \cite{Drummond:1980Prd, Shore:1996NP}.

\begin{acknowledgments}
The authors would like to thank I.~Lindell and M.W.~ McCall for numerous crucial discussions. AF is also grateful to G.M.~Thomas for suggesting a smoother wording to several concepts. This project was supported by the EPSRC, grant no.\ EP/E031463/1, and by the Academy of Finland, project no.\ 124204.
\end{acknowledgments}

\appendix
\section{Non-birefringent normal forms from Schuller's classification}
\label{sec:appA}

In this appendix, we classify all non-birefringent media by means typical representatives (normal matrix forms). There exist in total five normal forms (``NBR'' classes), which are obtained as limits from six specific ``metaclasses'' in Schuller's classification \cite{Schuller:2010Ap}. As an aside, one should observe that some of the results reported here also appear in Section \ref{sec:3D}. 

The first NBR class is encoded by the matrix
\begin{align}
\label{chii}
 \dd \chi_{\textindex{i}} = \begin{bmatrix}
             -\tau&0&0&\sigma&0&0\\
             0&-\tau&0&0&\sigma&0\\
             0&0&-\tau&0&0&\sigma\\
             \sigma&0&0&\tau&0&0\\
             0&\sigma&0&0&\tau&0\\
             0&0&\sigma&0&0&\tau
            \end{bmatrix}\ ,
\end{align}
which is obtained from metaclass I, with $\tau_1 = \tau_2 = \tau_3$ and $\sigma_1=\sigma_2=\sigma_3$. Equation \eqref{chii} covers all hyperbolic metric media, together with an axion term. The corresponding Fresnel equation is the free-space relation
\begin{equation}
 \frac{\omega^2}{c^2} = k_1^2 + k_2^2 + k_3^2\ .
\end{equation}

The second NBR class is attained from metaclass VII:
\begin{equation}
\label{chiii}
 \dd \chi_{\textindex{ii}} = \begin{bmatrix}
             \lambda_5&0&0&\lambda_6&0&0\\
             0&\lambda_3&0&0&\lambda_4&0\\
             0&0&\lambda_1&0&0&\lambda_2\\
             \lambda_6&0&0&\lambda_5&0&0\\
             0&\lambda_4&0&0&\lambda_3&0\\
             0&0&\lambda_2&0&0&\lambda_1
            \end{bmatrix}\ ,
\end{equation}
with the additional constraints
\begin{align}
 \lambda_1 &= s_1 (\lambda_2-\lambda_6) + s_2 \lambda_5\ , \\ \lambda_3 &= s_3 (\lambda_4-\lambda_6) + s_4 \lambda_5\ ,
\end{align}
where $s_1$, $s_2$, $s_3$ and $s_4$ are signs. The associated dispersion relation reads:
\begin{equation}
 (\frac{\omega^2}{c^2} + s_2 s_4 k_1^2 + s_2 k_2^2 + s_4 k_3^2)^2 = 0\ ,
\end{equation}
so that the optical metric has signature (4,0) when $s_{2}=s_{4}=+1$, and (2,2) otherwise. From equation \eqref{chiii} one can see that imposing $\lambda_2=\lambda_4=\lambda_6$ reduces this NBR class to (4,0) or (2,2) metric media -- with the addition of an axion piece. 

The remaining three NBR classes are ``degenerate'', in that pairs of metaclasses give the same non-birefringent limit. This can be explained as follows: $\sigma_i$ and $\tau_i$ are the real and imaginary parts of a complex eigenvalue $z_{i}$; the symbol $\lambda_i$ denotes a real eigenvalue. Consider, then, some NBR matrix, which is derived by setting $\tau_i = 0$ in a metaclass ``A''. Clearly, enforcing  $\lambda_j=0$ in a metaclass ``B'' with one complex eigenvalue less, yields the same ``degenerate'' result.

The third NBR class is given by 
\begin{equation}
\label{chiiii}
 \dd \chi_{\textindex{iii}} = \begin{bmatrix}
             0&0&0&\pm \lambda_1 + \lambda_2&0&0\\
             0&-\tau&0&0&\sigma&0\\
             0&0&\lambda_1&0&0&\lambda_2\\
             \pm \lambda_1 + \lambda_2&0&0&0&0&0\\
             0&\sigma&0&0&\tau&0\\
             0&0&\lambda_2&0&0&\lambda_1
            \end{bmatrix}\ .
\end{equation}
This matrix is obtained from metaclass IV with ($\tau_1 = 0$, $\sigma_1 = \pm \lambda_1 + \lambda_2$) or ($\tau_2 = 0$, $\sigma_2 = \pm \lambda_1 + \lambda_2$). If the second set of constraints is used, certain spacetime directions must be exchanged in order to retrieve Eq.\ \eqref{chiiii}. Alternatively, the above normal form is the limit ($\lambda_1 = 0$, $\lambda_2 = \pm \lambda_3 + \lambda_4$) or ($\lambda_3 = 0$, $\lambda_4 = \pm \lambda_1 + \lambda_2$) of metaclass VI.

The fourth NBR class has the representation
\begin{equation}
\label{chiiv}
 \dd \chi_{\textindex{iv}} = \begin{bmatrix}
             0&0&0&\pm \lambda_1 + \lambda_2&0&0\\
             0&\lambda_3&0&0&\lambda_4&0\\
             0&0&\lambda_1&&0&\lambda_2\\
             \pm \lambda_1 + \lambda_2&0&0&0&0&0\\
             0&\lambda_4&0&0&\lambda_3&0\\
             0&0&\lambda_2&0&0&\lambda_1
            \end{bmatrix}\ ,
\end{equation}
which is achieved from metaclass VI with ($\tau_1=0$, $\sigma_1 = \pm \lambda_1 + \lambda_2$), or from metaclass VII with ($\lambda_5=0$, $\lambda_6 = \pm \lambda_1 + \lambda_2$). Other possibilities do exist; however, they correspond to a mere relabelling of coordinates.

The fifth and last NBR class has the normal form
\begin{equation}
\label{chiv}
 \dd \chi_{\textindex{v}} = \begin{bmatrix}
             0&0&0&\lambda_1&0&0\\
             0&0&0&0&\lambda_5&0\\
             0&0&\pm(\lambda_3-\lambda_5)&0&0&\lambda_3\\
             \lambda_3&0&0&\epsilon_1&0&0\\
             0&\lambda_5&0&0&0&0\\
             0&0&\lambda_1&0&0&\pm(\lambda_3-\lambda_5)
            \end{bmatrix}\ ,
\end{equation}
where $\epsilon_1 = \pm 1$. Eq.\ \eqref{chiv} is attained from metaclass XVII by taking ($\tau_2=0$, $\lambda_2 = \pm(\lambda_3-\sigma_2)$). Equivalently, one can start from metaclass XVIII, and set ($\lambda_4=0$, $\lambda_2=\pm(\lambda_3-\lambda_5)$) or ($\lambda_2=0$, $\lambda_4=\pm(\lambda_3-\lambda_5)$).

The Fresnel equation of NBR classes iii, iv and v reads
\begin{equation}
 (\pm \frac{\omega}{c} k_3 + k_1 k_2)^2 = 0\ ,
\end{equation}
and corresponds to an optical metric with signature (2,2).

\section{The geometry of the NBR classes}
\label{sec:appB}
We shall now describe how to derive the general formula \eqref{allNBRchi} from the NBR matrices of Appendix \ref{sec:appA}. Given the dispersion relation of a normal form, the metric $\dd Q$ is fixed up to a constant. This specifies the ``core'' of the medium response: a ``remainder'' (or ``residue'') is left out, which is simplified  -- fully, or to some extent -- by a suitable choice of axion. At this point, the problem for NBR class i solved immediately, as one finds $\mbf A = \tilde{\mbf A} = 0$.

For NBR class ii, the residue reads:
\begin{equation}
\label{DeltaVII}
 \Delta \dd \chi_{\textindex{ii}} = \begin{bmatrix}
             0&0&0&0&0&0\\
             0&s_{\alpha_2} \alpha_2&0&0&\alpha_2&0\\
             0&0&s_{\alpha_1}  \alpha_1&0&0&\alpha_1\\
             0&0&0&0&0&0\\
             0&\alpha_2&0&0& s_{\alpha_2}\alpha_2&0\\
             0&0&\alpha_1&0&0&s_{\alpha_1} \alpha_1
            \end{bmatrix}\ ,
\end{equation}
where $s_{\alpha_2}$ and $s_{\alpha_1}$ are signs and satisfy $s_{\alpha_2} = s_2 s_4 s_{\alpha_1}$. By inspection, one can conclude that the structure \eqref{DeltaVII} does indeed consist of two bivectors $\mbf{A}$ and $\tilde{\mbf{A}}$.

For NBR classes iii, iv and v, the remainder is of the form (in the case of NBR class v, one must re-label the coordinates):
\begin{equation}
\label{DeltaIV}
  \Delta \dd \chi_{\textindex{iii/iv/v}} = \begin{bmatrix}
             0&0&0&0&0&0\\
             0&\alpha&0&0&\gamma&0\\
             0&0&0&0&0&0\\
             0&0&0&0&0&0\\
             0&\gamma&0&0&\beta&0\\
             0&0&0&0&0&0
            \end{bmatrix}\ .
\end{equation}
Again, a simple calculation reveals that this matrix can be represented by $\mbf{A}$ and $\tilde{\mbf{A}}$ (cf.\ Sec.\ \ref{sec:3D}). Consequently, Eq.\ \eqref{allNBRchi} \emph{must} encompass all non-birefringent media.  

As a final remark, explicit expressions for the bivectors can be deduced from the residues in this Appendix. The resulting quantities ($\mbf{A}$, $\mbf{\tilde{A}}$ and $\dd Q$) always comply with the conditions \eqref{ABconstrI}--\eqref{ABconstrII}. This constitutes an excellent consistency check. 

\section{ Schuller's classification and premetric electrodynamics}
\label{sec:appC}
The classification proposed by Schuller et al.\ \cite{Schuller:2010Ap} is (background) metric free by construction. Yet, it \emph{relies} on a linear medium response $\Omega^{\mu\nu\alpha\beta}$ which is a \emph{true tensor}:
\begin{equation}\label{transfomega}
  \Omega^{\mu'\nu'\alpha'\beta'}\!=L^{\mu'}_{\ \mu}L^{\nu'}_{\ \nu}L^{\alpha'}_{\ \alpha}L^{\beta'}_{\ \beta}\,\Omega^{\mu\nu\alpha\beta}\ .
\end{equation} 
In the premetric formalism of Sec.\ \ref{sec:3}, one must derive $\Omega^{\mu\nu\alpha\beta}$ from the canonical $\chi^{\mu\nu\alpha\beta}_{\textindex{PM}}$. For this purpose, Ref.\ \cite{Schuller:2010Ap} develops a scheme to eliminate the ``extra'' density factor in Eq.\ \eqref{chipmtransf}. The present appendix analyzes this procedure; it demonstrates its consistency, but reveals that a specific uniqueness proof is still missing. Equivalently, a (cautiously) positive answer is found to the question: can  Schuller's classification be employed in premetric electrodynamics?

Schuller et al.\ examine the quantity $\vert\det(\Omega^{IJ})\vert^{1/6}$, and show that it is a scalar density:
\begin{equation}\label{den6x6chisch}
  \vert\det(\Omega^{I'J'})\vert^{1/6}=\vert\det(L^{\rho'}_{\ \rho})\vert^{+1}\vert\det(\Omega^{IJ})\vert^{1/6}\ ,
\end{equation}
where $\det(\Omega^{IJ})$ is a 6$\times$6 determinant. As an aside, it is interesting to make a comparison with the true scalar $\vert\det(\chi_{\textindex{PM}}^{IJ})\vert$, satisfying:
\begin{equation}\label{scalar6x6det}
  \vert\det(\chi_{\textindex{PM}}^{I'J'})\vert=\vert\det(\chi_{\textindex{PM}}^{IJ})\vert\ .
\end{equation}
Resuming the derivation of Ref.\ \cite{Schuller:2010Ap}, the weight of $-1$ obtained in \eqref{den6x6chisch} is used to compensate the $w=+1$ scaling of $e^{\alpha\beta\gamma\delta}$ (see Eq.\ \eqref{densityperm}):    
\begin{equation}
  \upsilon^{\alpha\beta\gamma\delta}:=\vert\det(\Omega^{IJ})\vert^{1/6}e^{\alpha\beta\gamma\delta}.
\end{equation}
Remarkably, a Levi-Civita-type tensor, $\upsilon^{\alpha\beta\gamma\delta}$, is therefore achieved, which, under a change of frame, obeys
\begin{equation}
   \upsilon^{\alpha'\beta'\gamma'\delta'}=L^{\alpha'}_{\ \alpha}L^{\beta'}_{\ \beta}L^{\gamma'}_{\ \gamma}L^{\delta'}_{\ \delta}\, \upsilon^{\alpha\beta\gamma\delta}\ .
\end{equation}
A similar procedure is exploited to link $\Omega^{\mu\nu\alpha\beta}$ to the canonical $\chi_{\textindex{PM}}^{\mu\nu\alpha\beta}$, so as to attain the relation
\begin{equation}\label{schullerconsistent}
  \frac{\dd\chi_{\textindex{PM}}}{\vert\det(\chi_{\textindex{PM}}^{IJ})\vert^{1/6}}= \frac{\dd\Omega}{\vert\det(\Omega^{IJ})\vert^{1/6}}\ .
\end{equation}
Here, the expression on the right hand side can be found in \cite{Schuller:2010Ap}, while the expression on the left hand side is a slight generalization comprising those $\chi_{\textindex{PM}}^{\mu\nu\alpha\beta}$ that are not unimodular in a 6$\times$6 sense. Thus, combining the transformation properties (\ref{chipmtransf},\ref{transfomega},\ref{den6x6chisch},\ref{scalar6x6det}), one reaches the important conclusion that the density-elimination scheme proposed by Schuller et al.\ is fully consistent. 

Besides this positive outcome, one can observe that the two fractions in \eqref{schullerconsistent} are functions of two separate objects, $\Omega^{\mu\nu\alpha\beta}$ and $\chi^{\mu\nu\alpha\beta}_{\textindex{PM}}$ . Consequently, one can never \emph{uniquely} derive a tensorial $\Omega^{\mu\nu\alpha\beta}$ from the more fundamental $\chi_{\textindex{PM}}^{\mu\nu\alpha\beta}$. Accordingly, a density-free formulation, even if consistently constructed, does not appear to emerge unambiguously from the axiomatic structure of pre-metric electrodynamics \cite{Hehl:2003} -- at least so far. 

In this paper, there exists an intuitive way to account for densities: once a metric $Q^{\alpha\beta}$ is extracted from $\Omega^{\mu\nu\alpha\beta}$, the identity
\begin{equation}
\sqrt{\vert\det(Q^{-1}_{\alpha'\beta'})\vert}=\vert\det(L^{\rho'}_{\ \rho})\vert^{-1}\sqrt{\vert\det(Q^{-1}_{\alpha\beta})\vert}\ ,
\end{equation}
can be used. Consequently, one can readily make a transition to $\chi_{\textindex{PM}}^{\mu\nu\alpha\beta}$, as demonstrated in Eq.\ \eqref{allNBRchiPM}; the problem of uniqueness found in \eqref{schullerconsistent} is accounted by the prefactor $M$, and does not affect our results (Sec.\ \ref{sec:4A}).

\end{document}